\documentclass[%
 aip,
 amsmath,amssymb,
 reprint,%
]{revtex4-1}

\usepackage{graphicx}
\usepackage{dcolumn}
\usepackage{bm}

\usepackage[utf8]{inputenc}
\usepackage[T1]{fontenc}
\usepackage{mathptmx}

\begin{document}

\preprint{QPC reproducibility paper vAPL 3}

\title{Improving reproducibility of quantum devices with completely undoped architectures}

\author{A. Srinivasan}
\affiliation{School of Physics, University of New South Wales, Sydney NSW 2052, Australia}

\author{I. Farrer}
\affiliation{Department of electronic and electrical engineering, University of Sheffield, Sheffield S10 2TN, United Kingdom}

\author{D. A. Ritchie}
\affiliation{Cavendish Laboratory, University of Cambridge, Cambridge CB3 0HE, United Kingdom}

\author{A. R. Hamilton}%
\email{Alex.Hamilton@unsw.edu.au}
\affiliation{School of Physics, University of New South Wales, Sydney NSW 2052, Australia}

\date{\today}

\begin{abstract}
	The reproducible operation of quantum electronic devices is a key requirement for future quantum information processing and spintronics applications.  Traditionally quantum devices have been fabricated from modulation doped heterostructures, where there is an intrinsic lack of reproducibility due to the random potential from ionized donors.  Here we show that we can greatly improve reproducibility over modulation doped devices by using a completely undoped architecture, with superior uniformity in the confinement potential and more consistent operating voltages for both electron and hole devices.  Our results demonstrate that undoped heterostructures have significant advantages over modulation doping for reproducible manufacturing of quantum devices.
\end{abstract}

\maketitle


Quantum electronic devices such as quantum point contacts (QPCs) and quantum dots have generated significant research interest for applications in spintronics and quantum information processing, due to the potential for electrical spin control \cite{DattaApl90, LossPRA98, WolfSci01, AwsNPhys07, DebrayNNano09, ChenPRL12}.  These devices may form the building blocks of future quantum circuits, such as a qubit array based on a large number of identical quantum dots using QPCs as charge sensors. To achieve large scale manufacturability, it is critical to first establish reproducibility, such that each component in an integrated circuit has identical operating parameters.

Traditionally modulation doped structures have been used for quantum electronic devices due to the ease of fabrication.  However, the background electrostatic potential from randomly distributed ionized donors greatly reduces reproducibility \cite{YangAPL09,YakiJPCM16}. This intrinsic variability can be avoided by utilizing completely undoped structures, whereby the charge carriers are confined at a hetero-interface by applying the appropriate bias to a metallic top gate \cite{KaneAPL93, WilletAPL2006, LuAPL2009, ChenAPL12}.  These structures have many advantages including improved mobility \cite{MakAPL13}, improved thermal cycling properties \cite{SeePRL12}, and as we will show here, much superior reproducibility of quantum transport properties.

A quantum point contact, which is a narrow 1D channel linking two 2D reservoirs, is the simplest type of gate defined quantum device, making it ideal for the study of reproducibility \cite{YangAPL09, AltAPL13, LukePRB14}. We begin by asking the question: If several identical devices are made on the same wafer, do they exhibit the same behavior? To investigate this, we fabricate 18 nominally identical QPCs on modulation doped and undoped wafers, and look at the gate biases required to define and pinch off the 1D channels.  We also examine the conductance quantization and uniformity of the electrostatic potential inside the QPC channel, as well as reproducibility under thermal cycling.  

For comparison, we also investigate reproducibility in hole QPCs.  Hole quantum devices based on III-V semiconductor systems have attracted a lot of recent interest due to their complex spin properties arising from the strong bulk spin-orbit interaction in the valence band \cite{Winkler03}.  Hole systems therefore offer the ideal platform for electrical spin control, and could potentially be utilized for hole spin qubits for quantum computation \cite{DlossPRL07, PribNN13, HigginbothamNLETT2014, MaurandNComm16}.  Hence, establishing reproducibility in hole devices may be of equal importance to electron devices.

Starting with the electron devices, we fabricated several identical QPCs on a modulation doped (100) GaAs/AlGaAs wafer (W0191) and a completely undoped (100) GaAs/AlGaAs wafer (W805).  Fig. 1 shows a schematic of the two wafer structures. In both heterostructures, the two dimensional electron gas (2DEG) is formed at a GaAs/AlGaAs interface 90nm below the surface.  In the case of the undoped wafer, a 30nm oxide layer is deposited on the GaAs surface, followed by an overall topgate.  A positive bias ($V_{TG} = 1.20V$) is applied to the topgate, to induce the 2D electron system.
The 2DEG in these wafers have the same electron density  ($n = 1.8 \times 10^{11} cm^{-2}$), with mobilities of $\mu_{d} = 2 \times 10^6 cm^2V^{-1}s^{-1},~\mu_{u} = 4 \times 10^6 cm^2V^{-1}s^{-1}$ for the doped and undoped wafers respectively. The background impurity densities in the undoped wafer are $N_{b_{AlGaAs}} \approx 1.6 \times 10^{14} cm^{-3}$ and $N_{b_{GaAs}} \approx 6 \times 10^{13} cm^{-3}~$\cite{WangPRB13}.
The QPCs are defined by depositing split gates on the wafer surface, patterned using electron beam lithography.  The QPCs have lithographic dimensions 200nm long and 300nm wide (see SEM image in Fig. 1d).

From each wafer, three separate devices were fabricated, each of which had four QPCs, for a total of 12 QPCs.  Three of the QPCs on each wafer did not show characteristic 1D behavior, possibly due to fabrication errors, leaving 9 nominally identical QPCs for each wafer structure.  Electrical measurements were carried out at 4K in a liquid Helium cryostat, using standard ac lockin techniques.  The conductance of each QPC was measured as a function of split gate voltage $V_{SG}$ and compared to assess reproducibility.

\begin{figure}
	\includegraphics[width=0.99\linewidth]{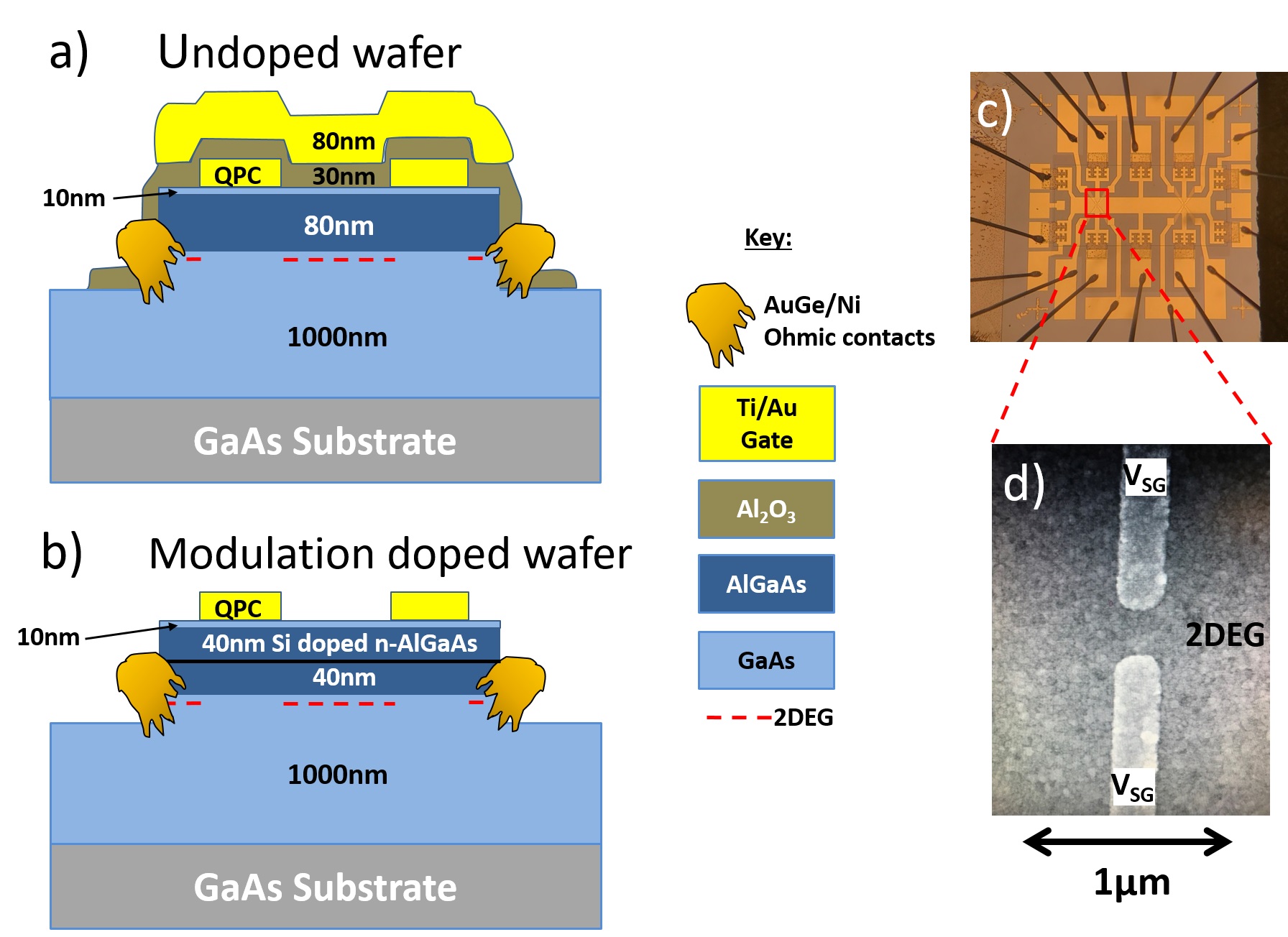}
	\caption{a) and b) show schematic diagrams of the wafer structures for the undoped and n-type modulation doped wafers respectively. c shows an optical image of the completed device on the undoped wafer, and d shows an electron microscope image of one of the QPCs.} 
\end{figure}

We now turn to the results.  Figs. 2a and 2b show conductance traces from 18 QPCs, 9 fabricated from the modulation doped wafer and 9 from the undoped wafer. Since the 2D electron density and the lithographic dimensions of the QPCs are identical for both wafers, both sets of QPCs show the same number of one dimensional modes (10) before reaching the definition point $V_d$, where there is a transition to the 2D regime.  However, it is evident that the undoped wafer shows noticeably less variation between different QPC traces.  

In order to compare and assess the reproducibility of both wafer types in detail, we look at two aspects.  Firstly, the operating characteristics (including pinch off voltage and definition point), and secondly, the conductance trace as a whole including the quantum transport features are compared.  


The inter-device reproducibility is quantified by calculating the spread of a parameter $x$, defined by:

\begin{equation}
\text{Spread} = \sigma_{x} / \overline{x}
\end{equation}

where $\sigma_{x}$ is the standard deviation and $\overline{x}$ is the mean.  We begin by looking at the conductance at which the 1D channel defines, $G_d$, since this is a measure of the electrical width of the channel.  Using eqn. 1, we find that spread of $G_d$ is 8.6$\%$ for the doped wafer, compared to a considerably improved 4.7$\%$ for the undoped wafer. 

To quantify the gate bias reproducibility, we look at variation of the pinch-off voltage $V_{po}$, and the definition voltage ${V_{d}}$.  Here, simply using the spread as defined in eqn. 1 is not appropriate, as the gate voltage range of the conductance traces are larger for the undoped wafer than the doped wafer due to the additional capacitance of the overall top gate.  Hence for comparing gate voltages, we instead use a \textit{relative} spread given by:
\begin{equation}
\text{ Relative spread}~(V_{po}) = \sigma_{V_{po}} / (\overline{V_{d}} - \overline{V_{po}}) 
\end{equation}
\begin{equation}
\text{ Relative spread}~(V_{d}) = \sigma_{V_{d}} / (\overline{V_{d}} - \overline{V_{po}}) 
\end{equation}

Using the 9 values of $V_{po}$ and ${V_{d}}$ for the doped wafer in eqn. 2, gives a relative spread for $V_{po}$ of 11.2$\%$. Repeating the calculation for the 9 values from the undoped wafer yields 6.7$\%$.   Similarly, using eqn. 3 we find a relative spread for ${V_{d}}$ of 2.3$\%$ for the doped wafer, and 1.7$\%$ for the undoped wafer.
The undoped wafer shows an improvement over the doped wafer in all three operating parameters.

\begin{figure}
	\includegraphics[width=0.99\linewidth]{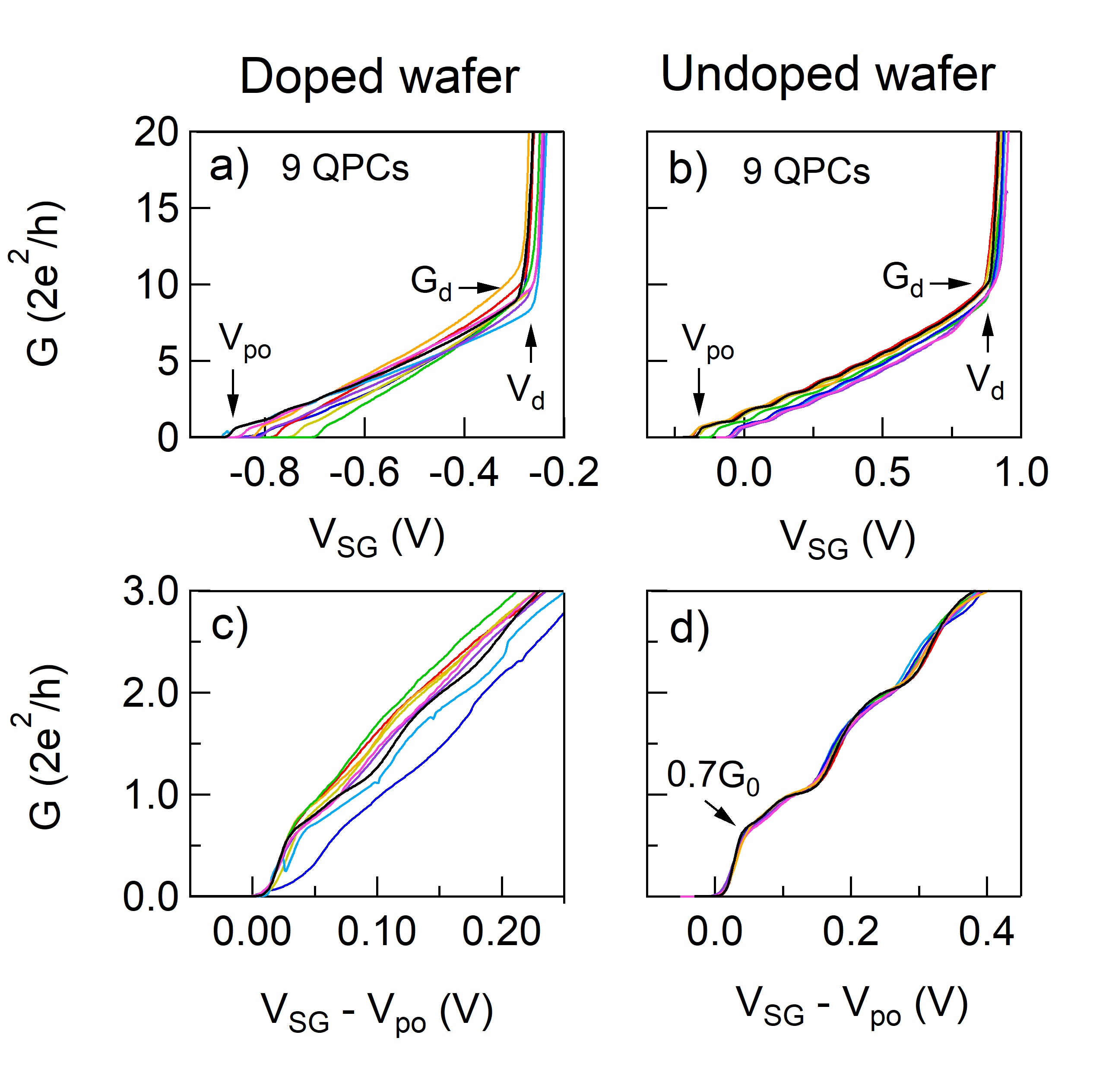}
	\caption{a) and b) show conductance traces of 18 QPCs from modulation doped and undoped wafers, with the operating parameters $V_{po}$, $G_d$ and $V_d$ labelled.  c) and d) show the conductance of the first few subbands after offsetting to remove the effect of variation in $V_{po}$.  The undoped wafer shows excellent reproducibility of quantum transport properties with all nine QPCs showing virtually identical conductance plateaus, risers and even 0.7 feature.}
\end{figure}

Now we compare the conductance trace as a whole, as well as the quantum transport features, which demonstrates the most significant advantage of undoped wafers over modulation doped wafers.  To begin, we consider the transconductance gain $\partial G/\partial V_{SG}$ which we take as the slope between $V_{po}$ and $V_d$.  We calculate the spread in $\partial G/\partial V_{SG}$ using eqn. 1 giving a value of 16$\%$ for the 9 modulation doped QPCs and only 5$\%$ for the undoped QPCs.  
This marked improvement in the reproducibility of the conductance is even more apparent in the quantum transport features of the lower 1D subbands. Figs. 2c and 2d show the conductance of the first few subbands for the modulation doped and undoped wafers respectively, where each trace for a given wafer has been offset in gate voltage to highlight variations in the conductance features.  The data from the modulation doped wafer in Fig. 2c do not show good reproducibility of the conductance features, with considerable variation in the plateau structures. This indicates that the potential inside each QPC is quite different despite the identical lithography. In contrast, all 9 conductance traces from the undoped wafer (fig. 2d) show virtually identical behavior, with the plateaus and risers between plateaus occurring at almost exactly the same gate voltage.  

Furthermore, the conductance quantization is clearly superior in the undoped wafer with the 9 QPCs showing between 4 and 7 visible plateaus at integer multiples of $2e^2/h$, compared to between 1 and 3 plateaus for the QPCs on the modulation doped wafer.  It is also significant that even the 0.7 anomaly \cite{ThomasPRL96} (indicated by the arrow in Fig. 2d) does not show any variability across the 9 undoped QPCs. It is known that the 0.7 feature is strongly dependent on the potential profile of the QPC \cite{ReillyPRB05, IqbalNat13, LukePRB14, LukePRB15}.  Despite this high sensitivity to the local potential, the 0.7 feature in all of the undoped QPCs are essentially identical, suggesting that the confinement potential in the undoped wafer must be extremely uniform.  This also confirms that small lithographic variations in the surface gates defining the QPCs have a negligible effect, and do not play a role in QPC variability.

We also investigate reproducibility under thermal cycling for both wafer structures.  Figs. 3a and 3b show the conductance traces of three QPCs on the modulation doped and undoped wafers measured on two successive cool downs. The solid and dotted lines show the traces obtained on the first and second cooldown respectively. The modulation doped QPCs show a random change after thermal cycling with $V_d$ and $V_{po}$ moving up and down in voltage for different QPCs, as shown by the black arrows in Fig. 3a. This is likely related to the random redistribution of ionised donors upon thermal cycling \cite{BuizPRL08,SeePRL12}.  In comparison, the QPCs on the undoped wafer show a smaller and more systematic movement, with all QPCs drifting in the same direction towards a more positive voltage after thermal cycling.

\begin{figure}
	\includegraphics[width=0.99\linewidth]{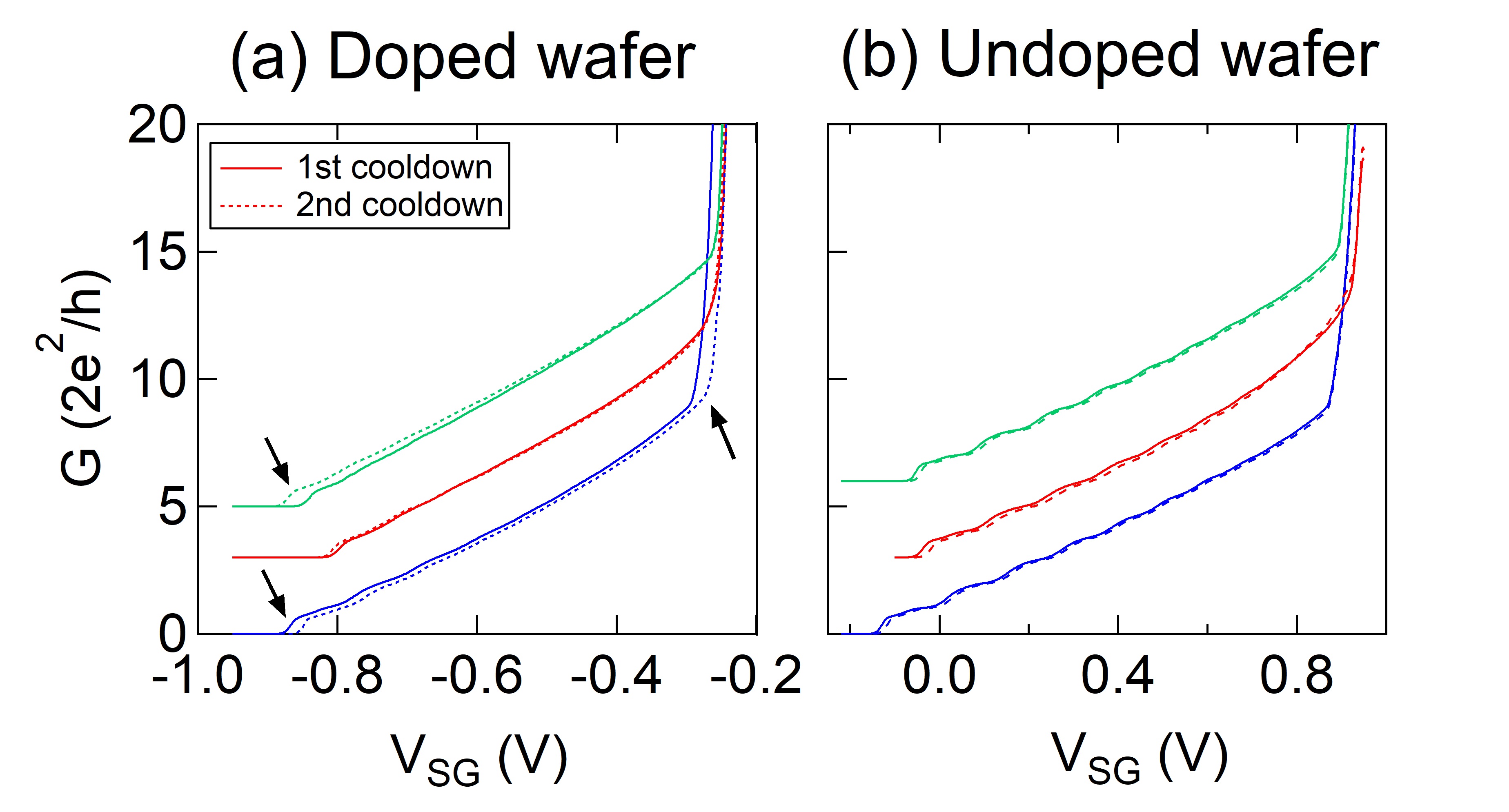}
	\caption{Conductance traces of three QPCs from the modulation doped wafer a) and the undoped wafer b) on two successive cool downs. The different QPCs have been offset vertically for clarity. The modulation doped wafer shows random variations upon thermal cycling (indicated by the black arrows), while the undoped wafer shows more consistent behavior.}
\end{figure}

Overall, compared to the modulation doped structure, the undoped structure shows much improved reproducibility of electron transport properties, with smaller variations in the operating voltages, superior thermal cycling properties, and conductance features that are extremely reproducible to a degree not possible with modulation doped structures.

We now show that the same principle can be extended to fabricating improved hole based devices.  The lack of reproducibility in modulation doped electron devices is exacerbated in modulation doped hole systems, making the fabrication of stable hole quantum devices difficult \cite{DaneshPRB97,RokhSNM02,GerlJCG07}.  
Here we demonstrate that by using the same undoped architecture shown in Fig. 1a, we can also fabricate hole QPCs with superior reproducibility to doped structures.  For comparison we also use a modulation doped hole system (W1071) with a similar structure to that shown in Fig. 1b, but with a carbon doped layer replacing the Si doped layer to achieve p-type doping.  Ohmic contact to the 2DHS is made using AuBe, instead of AuGe/Ni used for the electron systems. For the undoped wafers, a negative bias of -1.25V is applied to the topgate, to form a 2D hole system (2DHS) at the GaAs/AlGaAs interface. The 2DHS in these wafers have a 2D hole density of $p = 1.8 \times 10^{11} cm^{-2}$, and mobilities of $\mu_{d} = 2 \times 10^5 cm^2V^{-1}s^{-1},~\mu_{u} = 4 \times 10^5 cm^2V^{-1}s^{-1}$ for the doped and undoped wafers respectively. 
We again fabricated several identical QPCs on both the modulation doped and undoped wafers.  The hole QPCs were measured at T=250mK in a He$^3$ cryostat since 1D behavior cannot be observed at 4K for the hole devices (due to the smaller 1D subband spacing arising from the large hole effective mass).  

\begin{figure}
	\includegraphics[width=0.99\linewidth]{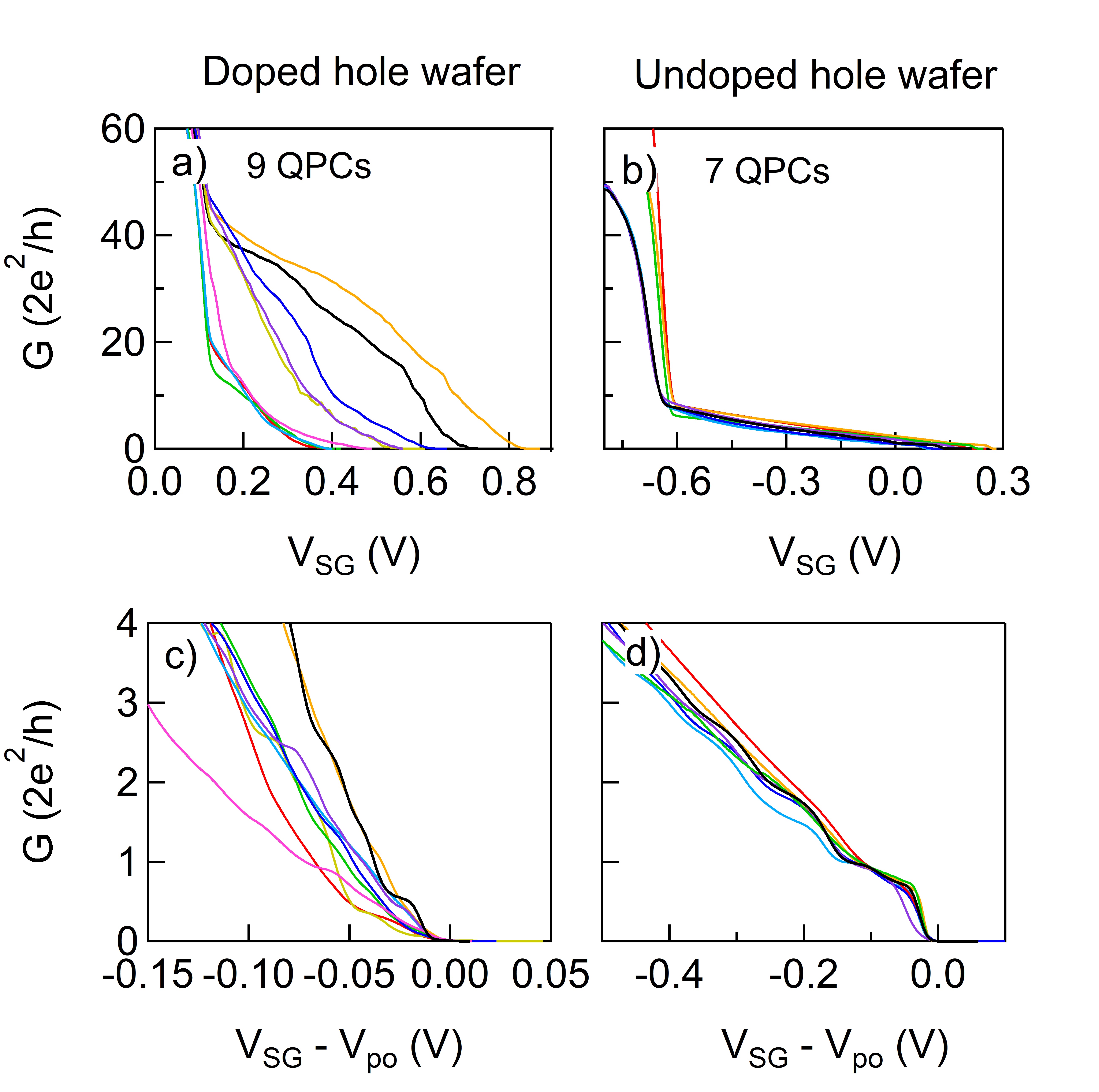}
	\caption{conductance traces of several hole QPCs from a) modulation doped and b) undoped wafers. c) and d) show the conductance of the first few subbands with the traces offset to the same pinch off voltage.  For hole systems, the disparity between modulation doped systems and undoped systems is enhanced further.}
\end{figure}

Fig. 4 shows conductance traces for (a) 9 modulation doped QPCs and (b) 7 undoped hole QPCs.  It is immediately evident that the modulation doped hole QPCs show extreme variability despite identical fabrication, while the undoped hole QPCs are far more reproducible.  We quantify the variability in the hole QPC parameters using the same method as for the electron QPCs.  Using eq. 1,
we find that the modulation doped hole QPCs have a large spread of $G_d$ of 44$\%$, versus 10.5$\%$ for the undoped hole QPCs.  Then using eq.2, the doped hole QPCs have a relative spread for $V_{po}$ of 37$\%$ versus only 8$\%$ for the undoped QPCs.  Fig. 5 summarises all the operating parameters for all electron and hole QPCs on the undoped and doped structures. 

\begin{figure}
	\includegraphics[width=0.99\linewidth]{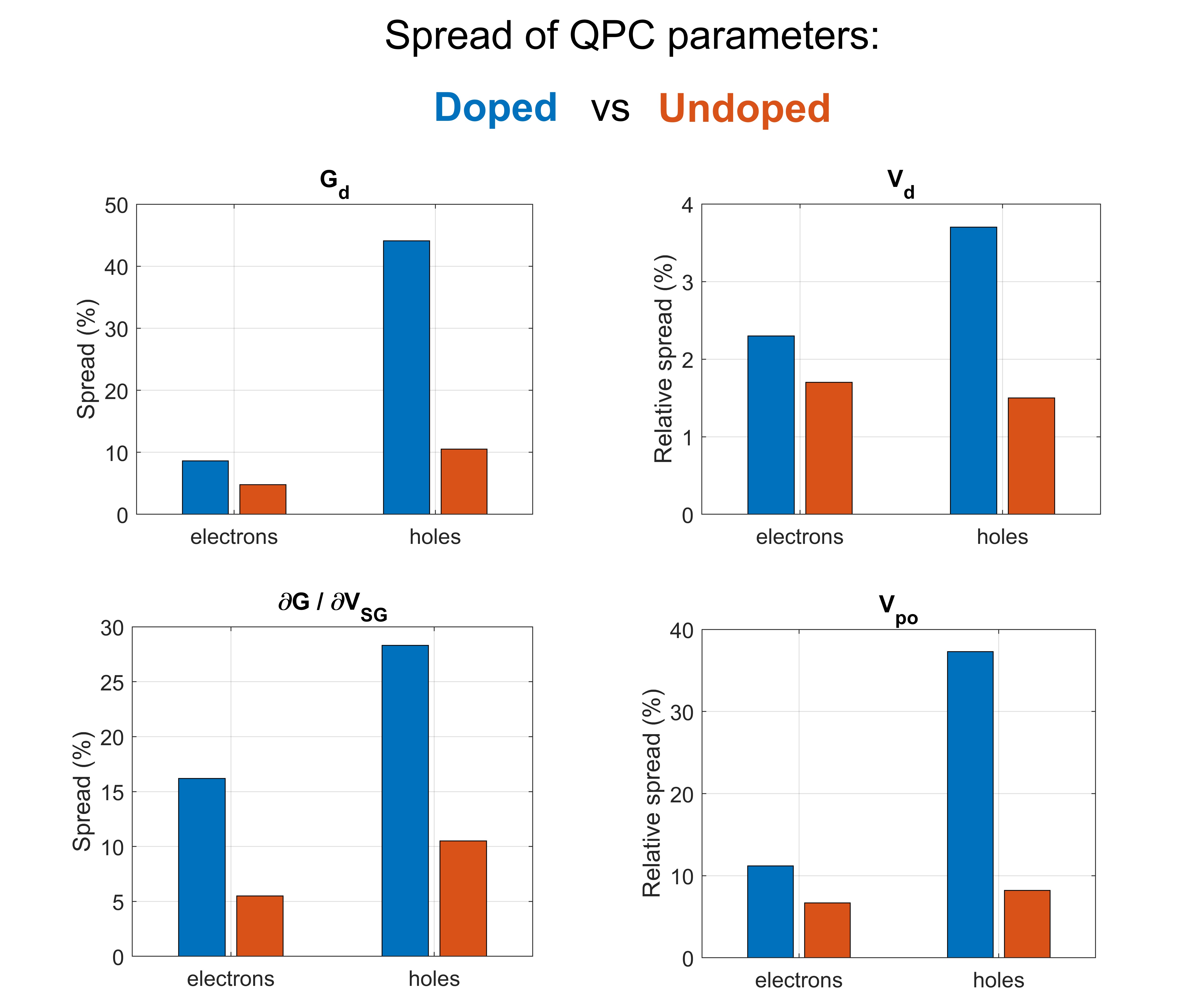}
	\caption{Summary of spread of operating parameters for the modulation doped and undoped electron and hole QPCs. Shorter bars are better. The values for $G_d$ and $\partial G/\partial V_{SG}$ are calculated using eqn. 1.  The values given for $V_d$, $V_{po}$ are a relative spread, calculated using eqns. 2 and 3. The undoped structures show superior reproducibility, particularly in the case of hole systems.}
\end{figure}

Furthermore, the modulation doped hole QPCs have drastic variability in conductance features, as shown in Fig. 4c, suggesting that the potential varies considerably between QPCs.  There is a large spread in the conductance slope $\partial G/\partial V_{SG}$ of 28$\%$ for the 9 modulation doped QPCs.  In the case of undoped holes (Fig. 4d), the overall conductance features are far more reproducible, with a spread of only 10.5$\%$ in the slope.  We also note that the quality of the conductance quantization is very different between the two types of devices.  Only 1 of 9 doped hole QPCs shows a plateau at $2e^2/h$ and none at $4e^2/h$, whereas 6 of the 7 undoped hole QPCs show a plateau at $2e^2/h$ and 4 of these also show a plateau at $4e^2/h$.

Finally, comparing the electron and hole QPCs on the undoped structure, we see that the electron QPCs have slightly improved reproducibility of operating parameters as shown in Fig. 5, and noticeably improved visibility and reproducibility of quantum transport features (Fig. 2d vs Fig. 4d).  This may be evidence that there are more background impurities ionized in a p-channel device than in an n-channel device, consistent with the findings of Chen et al.\cite{ChenAPL12}.  We also note that hole devices may be more sensitive to minor differences in potential due to the smaller energy scale as a result of the larger effective mass.

In conclusion, we have shown that quantum devices based on undoped heterostructures show significantly improved reproducibility compared to devices fabricated from modulation doped structures.  This is applicable to the design and manufacture of future quantum circuits which may require large numbers of identical quantum dots for qubit arrays.  These results are also useful for fabricating reproducible hole devices for studying the rich spin physics of hole systems.

\begin{acknowledgments}
This work was supported by the Australian Research Council under the Discovery Projects scheme, and was performed in part using facilities of the NSW Node of the Australian National Fabrication Facility.
\end{acknowledgments}

\section*{credit}
This article may be downloaded for personal use only. Any other use requires prior permission of the author and AIP Publishing. This article appeared in Appl. Phys. Lett. 117, 183101 (2020) and may be found at https://doi.org/10.1063/5.0024923

\section*{Data availability}
The data that support the findings of this study are available within the article.


\section*{References}
\begin{thebibliography}:
	
	\bibitem{DattaApl90} S. Datta and B. Das, Appl. Phys. Lett. {\bf 56}, 665 (1990).
	
	\bibitem{LossPRA98} D. Loss and D. P. DiVincenzo, Phys. Rev. A. \textbf{57}, 120 (1998).
	
	\bibitem{WolfSci01} S. A. Wolf, D. D. Awschalom, R. A. Buhrman, J. M. Daughton, S. von Molnar, M. L. Roukes, A. Y. Chtchelkanova, and D. M. Treger, Science {\bf 294}, 1488 (2001).
	
	\bibitem{AwsNPhys07} D. D. Awschalom, M. E. Flatte, Nat. Phys. \textbf{3}, 153-159 (2007).
	
	\bibitem{DebrayNNano09} P. Debray, S. M. S. Rahman, J. Wan, R. S. Newrock, M. Cahay, A. T. Ngo, S. E. Ulloa, S. T. Herbert, M. Muhammad, and M. Johnson, Nat. Nanotechnol. \textbf{4}, 759 (2009). 
	
	\bibitem{ChenPRL12} T. -M. Chen, M. Pepper, I. Farrer, G. A. C. Jones, and D. A. Ritchie, Phys. Rev. Lett. \textbf{109}, 177202 (2012). 
	
	\bibitem{YangAPL09} Q.-Z. Yang, M. J. Kelly, I. Farrer, H. E. Beere, and G. A. C. Jones, Appl. Phys. Lett. \textbf{94}, 033502 (2009). 
	
	\bibitem{YakiJPCM16} I. I. Yakimenko and K. F. Berggren, J. Phys.: Condens. Matter. \textbf{28}, 105801 (2016).
	
	\bibitem{KaneAPL93} B. E. Kane, L. N. Pfeiffer, and K. W. West, Appl. Phys. Lett. \textbf{63}, 2132 (1993).
	
	\bibitem{WilletAPL2006} R. L. Willett, L. N. Pfeiffer, and K. W. West, Appl. Phys. Lett. \textbf{89}, 242107 (2006).
	
	\bibitem{LuAPL2009} T. M. Lu, D. C. Tsui, C.-H. Lee, and C. W. Liu, Appl. Phys. Lett. \textbf{94}, 182102 (2009).
	
	\bibitem{ChenAPL12} J. C. H. Chen, D. Q. Wang, O. Klochan, A. P. Micolich, K. Das Gupta, F. Sfigakis, D. A. Ritchie, D. Reuter, A. D. Wieck, and A. R. Hamilton, Appl. Phys. Lett. \textbf{100}, 052101 (2012).
	
	\bibitem{MakAPL13} W. Y. Mak, F. Sgakis, K. Das Gupta, O. Klochan, H. E. Beere, I. Farrer, J. P. Griths, G. A. C. Jones, A. R. Hamilton, and D. A. Ritchie, Appl. Phys. Lett. \textbf{102}, 103507 (2013).
	
	\bibitem{SeePRL12} A. M. See, I. Pilgrim, B. C. Scannell, R. D. Montgomery, O. Klochan, A. M. Burke, M. Aagesen, P. E. Lindelof, I. Farrer, D. A. Ritchie, R. P. Taylor, A. R. Hamilton, and A. P. Micolich, Phys. Rev. Lett. \textbf{108}, 196807 (2012).
	
	\bibitem{AltAPL13} H. Al-Taie, L. W. Smith, B. Xu, P. See, J. P. Griffiths, H. E. Beere, G. A. C. Jones, D. A. Ritchie, M. J. Kelly, and C. G. Smith, Appl. Phys. Lett. \textbf{102}, 243102 (2013).
	
	\bibitem{LukePRB14} L. W. Smith, H. Al-Taie, F. Sfigakis, P. See, A. A. J. Lesage, B. Xu, J. P. Griffiths, H. E. Beere, G. A. C. Jones, D. A. Ritchie, M. J. Kelly, and C. G. Smith, Phys. Rev. B. \textbf{90}, 045426 (2014).
	
	\bibitem{Winkler03} R. Winkler, \textit{Spin-orbit coupling effects in two-dimensional electron and hole systems}, (Springer Tracts in Modern Physics, Vol. 191, Springer, Berlin, 2003).
	
	\bibitem{DlossPRL07} D. V. Bulaev and D. Loss, Phys. Rev. Lett. {\bf 98}, 097202 (2007).
	
	\bibitem{PribNN13} V. S. Pribiag, S. Nadj-Perge, S. M. Frolov, J. W. G. van den Berg, I. van Weperen, S. R. Plissard, E. P. A. M. Bakkers and L. P. Kouwenhoven, Nat. Nano, \textbf{8}, 170 (2013).
	
	\bibitem{HigginbothamNLETT2014} A. P. Higginbotham, T. W. Larsen, J. Yao, H. Yan, C. M. Lieber, C. M. Marcus, and F. Kuemmeth, Nano Lett. \textbf{14},  3582 (2014).
	
	\bibitem{MaurandNComm16} R. Maurand, X. Jehl, D. Kotekar-Patil, A. Corna, H. Bohuslavskyi, R. Lavieville, L. Hutin, S. Barraud, M. Vinet, M. Sanquer and S. De Franceschi, Nat. Comms. \textbf{7}, 13575 (2016).
	
	\bibitem{WangPRB13} D. Q. Wang, J. C. H. Chen, O. Klochan, K. Das Gupta, D. Reuter, A. D. Wieck, D. A. Ritchie, and A. R. Hamilton, Phys. Rev. B, \textbf{87}, 195313 (2013).
	
	\bibitem{ThomasPRL96} K. J. Thomas, J. T. Nicholls, M. Y. Simmons, M. Pepper, D. R. Mace, and D. A. Ritchie, Phys. Rev. Lett. \textbf{77}, 135 (1996).
	
	\bibitem{ReillyPRB05} D. J. Reilly, Phys. Rev. B \textbf{72}, 033309 (2005).
	
	\bibitem{IqbalNat13} M. J. Iqbal, R. Levy, E. J. Koop, J. B. Dekker, J. P. de Jong, J. H. M. van der Velde, D. Reuter, A. D. Wieck, R. Aguado, Y. Meir, and C. H. van der Wal, Nature (London) \textbf{501}, 79 (2013).
	
	\bibitem{LukePRB15} L. W. Smith, H. Al-Taie, A. A. J. Lesage, F. Sfigakis, P. See, J. P. Griffiths, H. E. Beere, G. A. C. Jones, D. A. Ritchie, A. R. Hamilton, M. J. Kelly, and C. G. Smith, Phys. Rev. B \textbf{91}, 235402 (2015).
	
	\bibitem{BuizPRL08} C. Buizert, F. H. L. Koppens, M. Pioro-Ladrie`re, H.-P. Tranitz, I.T. Vink, S. Tarucha, W. Wegscheider, and L. M. K. Vandersypen, Phys. Rev. Lett. \textbf{101}, 226603 (2008).
	
	
	\bibitem{DaneshPRB97} A. J. Daneshvar, C. J. B. Ford, A. R. Hamilton, M. Y. Simmons, M. Pepper, and D. A. Ritchie, Phys. Rev. B \textbf{55}, 13409 (1997).
	
	\bibitem{RokhSNM02} L. P. Rokhinson, D. C. Tsui, L. N. Pfeiffer, and K. W. West, Superlatt. Microstruct. \textbf{32}, 99 (2002).
	
	\bibitem{GerlJCG07} C. Gerl, J. Bauer, and W. Wegscheider, J. Cryst. Growth \textbf{301}, 145 (2007).

	
\end{thebibliography}
\end{document}